\documentclass[9pt,conference]{DAC}

\usepackage{cite}
\usepackage{amsmath,amssymb,amsfonts}
\usepackage{algorithmicx}
\usepackage{graphicx}
\usepackage{textcomp}
\usepackage{xcolor}
\usepackage{pifont}
\usepackage{array}
\usepackage{amsmath} 
\usepackage{bigstrut}
\usepackage[square,sort,comma,numbers,compress]{natbib}
\usepackage{algorithm} 
\usepackage{algpseudocode}
\usepackage{makecell}
\usepackage{bbding}
\usepackage{flushend}
\usepackage{bm}
\usepackage{hyperref}       
\hypersetup{
    colorlinks=true,
    linkcolor=blue,
    citecolor=blue,
    filecolor=magenta,      
    urlcolor=cyan,
    pdftitle={Overleaf Example},
    pdfpagemode=FullScreen,
    }

\usepackage{tabularx}
\usepackage{threeparttable}
\usepackage{xspace}
\def\BibTeX{{\rm B\kern-.05em{\sc i\kern-.025em b}\kern-.08em
    T\kern-.1667em\lower.7ex\hbox{E}\kern-.125emX}}

\usepackage{multirow}

\begin{document}

\title{Battle Against Fluctuating Quantum Noise: Compression-Aided Framework to Enable \\Robust Quantum Neural Network\\
{
\centering
\author{
\setlength{\baselineskip}{20pt}
Zhirui Hu$^{1,2}$, Youzuo Lin$^{3}$, Qiang Guan$^{4}$, Weiwen Jiang$^{1,2}$
\\
$^{1}$Electrical and Computer Engineering Department, George Mason University, Fairfax, VA 22030, US\\
$^{2}$Quantum Science and Engineering Center, George Mason University, Fairfax, VA 22030, US 
\\$^{3}$Earth and Environmental Sciences Division, Los Alamos National Laboratory, NM, 87545, US
\\ $^{4}$Department of Computer Science, Kent State University, 800 E Summit St, Kent, OH 44240
\\ (zhu2@gmu.edu; wjiang8@gmu.edu)
}
}
}


\maketitle

\begin{abstract}
Recently, we have been witnessing the scale-up of superconducting quantum computers; however, the noise of quantum bits (qubits) is still an obstacle for real-world applications to leveraging the power of quantum computing.
Although there exist error mitigation or error-aware designs for quantum applications, the inherent fluctuation of noise (a.k.a., instability) can easily collapse the performance of error-aware designs.
What's worse, users can even not be aware of the performance degradation caused by the change in noise.
To address both issues, in this paper we use Quantum Neural Network (QNN) as a vehicle to present a novel compression-aided framework, namely \textit{QuCAD}, which will adapt a trained QNN to fluctuating quantum noise.
In addition, with the historical calibration (noise) data,
our framework will build a model repository offline, which will significantly reduce the optimization time in the online adaption process.
Emulation results on an earthquake detection dataset show that QuCAD can achieve 14.91\% accuracy gain on average in 146 days over a noise-aware training approach. For the execution on a 7-qubit IBM quantum processor, ibm-jakarta, QuCAD can consistently achieve 12.52\% accuracy gain on earthquake detection.

\end{abstract}


\section{Introduction}

\begin{figure}[t]
\centering
\includegraphics[width=0.9\linewidth]{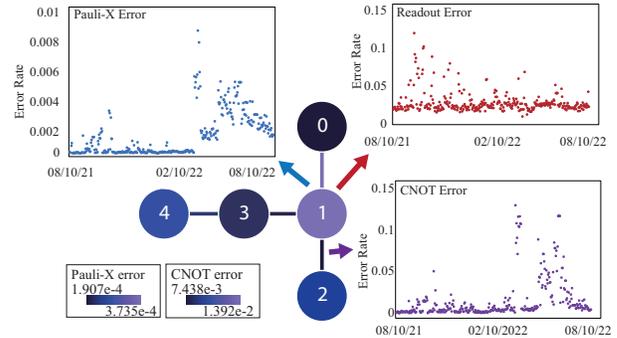}
\caption{The fluctuating noise observed on IBM backend belem}
\label{fig:introduction}
\end{figure}

We are currently in the Noisy Intermediate-Scale Quantum (NISQ) era where noise and scalability have been two well-known and critical issues in quantum computing. 
Nowadays, we have been witnessing the rapid development of superconducting quantum computers and the scalability issue is gradually mitigated. As an example, IBM took only 6 years to scale up its quantum computers from 5 to 433 qubits and the development of quantum research\cite{liang2022variational,liang2021can,wang2021exploration,jiang2021co,hu2022design} is advancing rapidly.
However, the high noise in quantum computing is still an obstacle for real-world applications to take advantage of quantum computing.
There are many sources of noise in quantum computing, such as gate errors and readout errors.
As shown in Fig.~\ref{fig:introduction}, the color of qubits and their connections indicates the Pauli-X and CNOT gate error from the IBM Belem backend.

Unlike CMOS noise that is within a small range under $10^{-15}$ so classical computing is more focused on performance efficiency rather than noise\cite{yang2023device,yang2022hardware,liao2021shadow}, the noise on qubits can reach $10^{-2}$ to $10^{-4}$.
Moreover, as shown in Fig.~\ref{fig:introduction}, noise from 1-year-long profiling on IBM Belem backend is fluctuating in a wide range, called ``fluctuating quantum noise''.
Although there exist works to improve the robustness of quantum circuits to noise, such as error mitigation \cite{takagi2022fundamental} and noise-aware designs \cite{ji2022calibration,bhattacharjee2019muqut,wang2022quantumnat}, these works commonly perform the optimization based on the noise at one moment.
The fluctuating quantum noise can easily make the quantum circuit lose its robustness.
Thus, new innovations are needed to deal with the fluctuating noise.

In this paper, we propose a novel framework, namely ``QuCAD'', to address the above issues. 
To illustrate our framework, we use Quantum Neural Network (QNN) --- a.k.a, variational quantum circuit (VQC) as an example --- since the learning approach has been shown to be an effective way for a wide range of applications from different domains (such as chemistry \cite{sajjan2022quantum}, healthcare \cite{maheshwari2022quantum}, and finance \cite{orus2019quantum}); and meanwhile, \cite{liu2021rigorous} recently has shown the potential quantum speedups using VQC.
To deal with fluctuating quantum noise, a straightforward method is to apply a noise-aware training approach to retrain QNN before each inference; however, it will obviously incur high costs.
More importantly, as the quantum noise changes in a wide range, we observe that a set of parameters will deviate from the loss surface, which impedes the noise-aware training to find optimal solutions.

To address this problem, we made an investigation and found that these parameters will lead to short circuits after the logical-to-physical compilation.
Equivalent to classical neural networks, these QNN parameters can be regarded as compression levels.
Therefore, QNN compression can be helpful to optimize the model in a noisy environment. 
Based on our observations, in QuCAD, we develop a novel noise-aware QNN compression algorithm.
It will interplay with other two components: (1) a model repository constructor that can build a repository of compressed models offline using representative historical data; and (2) a model repository manager that can make an online adaption of models to the fluctuating noise.
As such, QuCAD can automatically and efficiently obtain the QNN models adapted to run-time noise for high performance.

The main contributions of this paper are as follows.

\begin{itemize}
    \item We reveal that the fluctuating quantum noise will collapse the performance of quantum neural networks (QNNs).
    \item We develop a noise-aware QNN compression algorithm to adapt pretrained QNN model to a given noise.
    \item On top of the noise-aware compression algorithm, we further propose a 2-stage framework to adapt QNN model to fluctuating quantum noise automatically. 
    
\end{itemize}




Evaluation is conducted on both IBM Aer simulator and IBM quantum processor ibm-jakarta.
Experiment results on MNIST, earthquake detection dataset, and Iris show the effectiveness and efficiency of QuCAD. Especially, QuCAD achieves superior performance among the competitors on all datasets.
It achieves 16.32\%, 38.88\%, and 15.36\% accuracy gain compared with the default configuration without optimization. 
Even compared with noise-aware training before every execution, QuCAD can still achieve 
over a 15\% accuracy improvement on average; meanwhile, QuCAD can reduce the training time to over 100$\times$ at the online stage. 
For the execution on a 7-qubit IBM quantum processor ibm-jakarta, QuCAD can consistently achieve a 13.7\% accuracy gain on the earthquake detection dataset.

The remainder of the paper is organized as follows. Section 2
reviews the related work and provides observations, challenges and motivations. Section 3 presents the proposed
QuCAD framework. Experimental results are provided in Section 4 and concluding remarks are given in Section 5.



\section{Related Work, Challenge, and Motivation}



\subsection{Related Work}

There are two commonly applied approaches to mitigate the effect of quantum noise. One is to adjust quantum circuits by noise-aware training \cite{wang2022quantumnat} and mapping \cite{bhattacharjee2019muqut}. Noise-aware training is an adversary method, which injects noise into the training process so that the parameter of model can be learned from both datasets and devices. Noise-aware mapping is to minimize the sum of accumulated noise by changing the mapping from logical qubits to physical qubits. Another method is to estimate the ideal outputs by post-processing measurement data, e.g., zero-error noise extrapolation \cite{giurgica2020digital}, probabilistic error cancellation\cite{endo2018practical}, and virtual distillation\cite{huggins2021virtual}. However, these two kinds of approaches are based on noise at one moment, so we can hardly maintain the performance without redoing these methods as noise changes anytime.

There are a few works addressing the fluctuating-noise issue, which can be quantitatively analyzed in two aspects: the noise of device and the distance between ideal results and noisy results. \cite{dasgupta2021stability} evaluated the stability of NISQ devices with multiple metrics that characterize stability. \cite{dasgupta2022characterizing,dasgupta2022assessing} defined Hellinger distance, computational accuracy and result reproducibility to expound the distance between ideal result and noisy observation. These works show the significance of reproducing results in a long-term execution on a quantum device. However, it is still not clear what effects the fluctuating noise will make on quantum applications.


\subsection{Observation, Challenge and Motivation}
This section will provide our observations on the simulation performance of a quantum neural network model on a noisy quantum computer over a period of one year.

\begin{figure}[t]
\centering
\includegraphics[width=1\linewidth]{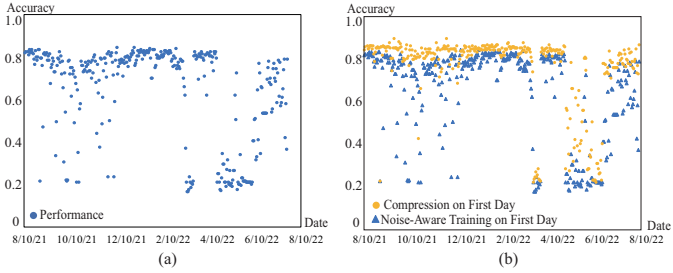}
\caption{The accuracy of QNN on 4-class MNIST from August 2021 to August 2022 on IBM backend belem using Qiskit Simulation.}
\label{fig:motivation1}
\end{figure}

\textbf{Observation 1: Fluctuating noise can collapse the model accuracy of a noise-aware trained QNN model.}

Fig.~\ref{fig:motivation1} (a) shows the daily accuracy of a QNN model on the quantum processor in Fig.~\ref{fig:introduction}, which demonstrates that the accuracy can be varied significantly along with the fluctuating noise.
The QNN model is trained using the noise-aware training method \cite{,wang2022quantumnat} based on the calibration (noise) data on day 1 (8/10/22).
Its accuracy is over 80\% from day 1 to day 22; however, when error rates increased on day 24, the accuracy decreased to 22\%.
This observation shows that the noise-aware design can obtain a robust quantum circuit to a certain range of noise, but the performance can be collapsed when the noise rate goes beyond a threshold.


 

\begin{figure}[t]
\centering
\includegraphics[width=1\linewidth]{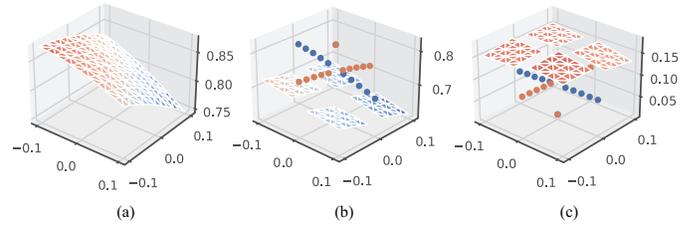}
\caption{Noise-aware training may miss optimal solution: (a) Optimization surface of 2-parameter VQC under noise free environment. (b) Optimization surface of the same VQC under a noisy environment. (c) Difference between (a) and (b).}
\label{fig:loss_3d}
\end{figure}

\textbf{Challenge 1: Noise on qubits can create breakpoints beyond the optimization surface, creating difficulties in training}


Relying on training only to adapt to noise may miss the opportunity to find the optimal solution.
Fig. \ref{fig:loss_3d}(a)-(b) shows an example of the results of a quantum circuit with 2 parameters (x-axis and y-axis) in a perfect environment (simulation) and in a noisy environment (realistic).
From the difference shown in Fig. \ref{fig:loss_3d}(c), we can see that the breakpoints in the loss landscape are with much lower noise.
Given the optimization started from a random point, in order to reach these breakpoints, it requires a dedicated learning rate for each parameter, which is almost impossible in a training algorithm.

\textbf{Motivation 1: Not only relying on training but also involving compression to battle against quantum noise.}

Looking closer at these breakpoints in Fig. \ref{fig:loss_3d}(c), we observe that the parameter value of 0 creates these breakpoints. We also observe such breakpoints around other specific values, like $\frac{\pi}{2}$, $\pi$, $\frac{3\pi}{2}$, etc.
We further investigate the root cause and we found the improvement in performance is caused by the reduction of physical circuit length.
More specifically, due to the transpilation (i.e., logical-to-physic qubit mapping), the logical quantum gate with different values will result in varied circuit lengths on physical qubits.
Analog to the classical neural networks, the reduction of the network complexity is known as compression.
Based on the above understanding, it motivates us to employ compression techniques to address the quantum noise.
For the same noise setting as Fig. \ref{fig:motivation1}(a), we perform the model compression on Day 1, and results are reported in Fig. \ref{fig:motivation1}(b).
It is clear that the results of compression (i.e., yellow points) are much better than the noise-aware training (i.e., blue points).
However, we still observe a significant accuracy drop between March 15 to May 29.


\textbf{Observation 2: Behaviors of fluctuating noise on different qubits has heterogeneity.}

\begin{figure}[t]
\centering
\includegraphics[width=1\linewidth]{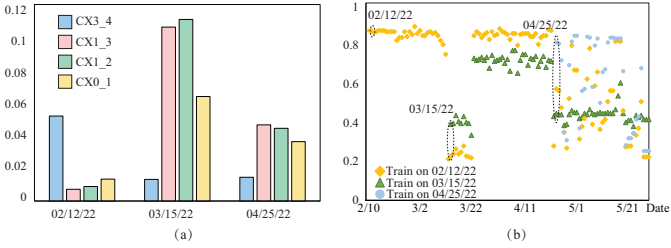}
\caption{Noise-aware compression is needed: (a) CNOT gate noise on three days. (b) Noise-aware compressing and tuning models on the three days in (a) and testing the on the following days.
 }
\label{fig:motivation2}
\end{figure}

To explore the root cause of the accuracy drop, we investigate the change of CNOT noise on all pairs of qubits on three dates: (1) Feb. 12, (2) March 15, and (3) April 25.
Fig.\ref{fig:motivation2} (a) reports the results.
It is clear that the qubits on March 15 and April 25 have much higher noise than that on Feb. 12.
A more interesting observation is that the fluctuating noise on different qubits has heterogeneity.
Specifically, on Feb. 12, $\langle q_3,q_4\rangle$ has the highest noise, while $\langle q_1,q_2\rangle$ becomes the noisiest one on March 15 and April 25.
Along with the heterogeneous changes of noise on qubits, the compressed QNN model may lose its robustness, which causes the accuracy degradation between March 15 to May 29.



\textbf{Motivation 2: Adding noise awareness in compression to battle against fluctuating quantum noise}


To address the above challenges, only conducting noise-agnostic compression with the objective of minimizing circuit length is not enough; on the other hand, we need to take the heterogeneous noise on qubits into consideration.
This motivates us to develop a ``noise-aware compression'' on QNNs, which can involve the noise level at each qubit to guide how the model will be compressed. Details will be introduced in Sec \ref{sec:framework}.
We use an example to illustrate the necessity of noise-aware compression. As shown in Fig. \ref{fig:motivation2} (b), on March 15 and April 25, we observe significant changes in the noise of qubits, which makes the original compressed model suffer an accuracy drop from 79\% to 22.5\% and 56.5\%.
By using a noise-aware compression on these dates, we resume the accuracy to 38.5\% and 80\% on March 15 and April 25, respectively.

With noise-aware compression, now, the problem is how frequently should we perform the compression.
Due to the fluctuation of quantum noise, one straightforward idea is to leverage the noise-aware compression before using it, so that it can adapt to the new noise. However, this can be too costly.



\textbf{Observation 3: Models can be re-utilized.}

Let's see the original model in Fig. \ref{fig:motivation2}(b), which experienced a dramatic accuracy drop on March 15.
However, after 9 days, on March 24, the accuracy of this model resumed to 80.5\%. Therefore, the previous model can be re-utilized later.

\textbf{Motivation 3: A model repository can avoid model optimization every day to improve efficiency.}

The above observation inspires us to build a model repository to keep a set of models.
At run time, instead of directly performing optimization (i.e., compression) for a new noise, we can first check if models in the repository can adapt to the noise.
In this way, it is possible to reuse pre-optimized models and significantly reduce the optimization cost at run time.

With the above observations and motivations, we propose a novel framework QuCAD in the following section to devise a noise-aware compression algorithm, build a model repository upon historical data offline, and manage it at run time.

\section{Compression-Aided Framework}

This section will introduce our proposed compression-aided framework, namely QuCAD, to battle against fluctuating quantum noise.
Before introducing details, we first formally formulate the problem as follows:
Given a QNN model $M$, the quantum processor $Q$ with history calibration (noise) data $D_t$, the current calibration data $D_c$ of $Q$, the problem is how to leverage the calibration data (i.e., $D_t$ and $D_c$) to find a model $M^{\prime}$ with the objective of maximizing the accuracy of $M^{\prime}$ on $Q$ with $D_c$.


\subsection{Framework Overview}\label{sec:framework}

\begin{figure}[t]
\centering
\includegraphics[width=0.9\linewidth]{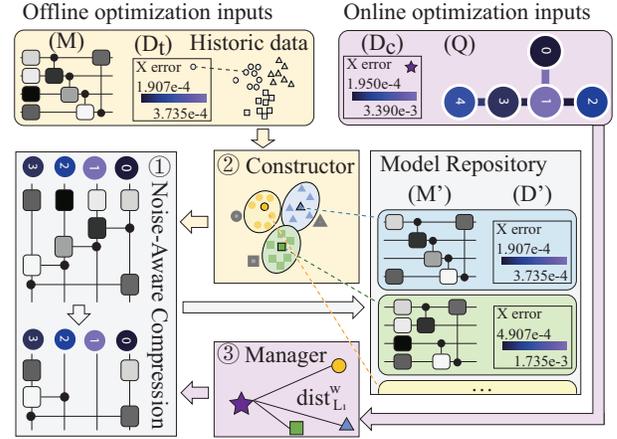}
\caption{Illustration of the proposed Compression-Aided Framework (QuCAD).}
\label{fig:overview}
\end{figure}

Fig.~\ref{fig:overview} shows the overview of the QuCAD framework.
It contains 3 main components: (1) a noise-aware compression algorithm, (2) an offline model repository constructor, and (3) an online model repository manager.
The noise-aware compression algorithm is the core of QuCAD in both offline and online optimizations.

The offline optimization includes the following steps to build a model repository.
We will first use the historical calibration data to obtain their corresponding performance for the given QNN model $M$.
Then, a clustering algorithm is developed to create several groups in terms of calibration data and corresponding model performance.
The centroid of calibration data in each group ($D^{\prime}$) will be used to optimize model $M$ the compression algorithm to generate the compressed model $M^{\prime}$.
The pair of $\langle M^{\prime}, D^{\prime}\rangle$ will be placed 
into the model repository.


The online optimization will get the current calibration data $D_c$ of quantum computer $Q$ as input. 
We match $D_c$ with the existing calibration data $D^{\prime}$ of items $\langle M^{\prime}, D^{\prime}\rangle$ in the model repository, aiming at finding the most similar one.
The model repository manager will use the difference between calibration data $D_c$ and $D^{\prime}$ to make a judgment whether model $M^{\prime}$ can be directly used under $D_c$.
If the distance is over a pre-set threshold then performance degradation is predicted,  
we will regard the current calibration data as a new centroid and generate a new model by noise-aware compression and put it into the model repository. 
Otherwise, we will output matched model directly. 


\subsection{Noise-Aware Compression}


Quantum neural network compression was recently proposed \cite{hu2022quantum} to reduce circuit length, where an Alternating direction method of multipliers (ADMM) approach is applied.
In this paper, we will also employ ADMM as the optimizer for noise-aware compression, and the new challenge here is how to involve noise awareness in the compression process.
In the previous work, the authors conducted compression upon the logical quantum circuit.
However, to involve noise in the compression of quantum gates, we need to fix the physic qubits of each quantum gate.
\textit{Therefore, we will take the quantum circuit after routing on restricted topology as input, instead of the logical quantum circuit.}

Before introducing our proposed algorithm, we first define the notations and the optimization problem which will be used in ADMM.
We denote $\bm{T}$ as a table of compression-level, which are breakpoints in the example of \textit{Motivation 1}.
We denote the function of a VQC under a noise-free (a.k.a., perfect, denoted by $p$) environment as $W_p(\bm{\theta})$, where $\bm{\theta} = [\theta_1,\theta_2,\cdots,\theta_n]$ are a set of trainable parameters, and $\theta_i$ is the parameter of gate $g_i$.
With the consideration of noise, the function of VQC will be changed to $W_n(\bm{\theta})$. 
On top of these, the deviation caused by noise can be defined as $N(\bm{\theta}) = W_n(\bm{\theta}) - W_p(\bm{\theta})$.
For a quantum gate $g_i$, it is associated with a physic qubit $q_k$ or a pair of physic qubits $\langle q_k,q_l\rangle$.
For simplicity, we denote such an association as a function $\langle q_k,q_l\rangle=A(g_i)$, and we use $k=l$ to represent $g_i$ is associated with one qubit $q_i$.
We denote $\bm{C}$ as a table of calibration data, and the notation $n_{k,l}\in \bm{C}$ or the function $C(q_k,q_l)$ to represent the noise rate on qubit $q_k$ and $q_l$ (or $q_k$ if $k=l$).
Then, the problem can be formulated as below.
\begin{equation}
\begin{aligned}
  &\min_{\bm{\theta}} \quad W_p(\bm{\theta}) + N(\bm{\theta})
\end{aligned}
\label{eq:optprob}
\end{equation}
Now, to enable using ADMM to 
perform noise-aware compression,
we first decompose the optimization problem in Eq. \ref{eq:optprob} to two sub-problems and solve them separately: (1) maximize accuracy and (2) minimize the deviation caused by noise.
The first subproblem can be solved by a gradient-based optimizer. For the second problem, we will use a set of auxiliary variables and an indicator function to resolve it, which will be introduced later.
Therefore, we reformulate the optimization problem that can be solved by ADMM as below.
\begin{equation}
\begin{aligned}
  &\min_{\{ \theta_i\}} \quad f(W_p(\bm{\theta})) + N(\bm{Z}) + \sum_{\forall z_i\in\bm{Z}}{s_i(z_i)},  \\
\end{aligned}
\label{eq:admm_train}
\end{equation}
where $\bm{Z}$ is a set of auxiliary variables for subproblem decomposition and $z_i\in \bm{Z}$ is corresponding to $\theta_i\in \bm{\theta}$; function $f$ represents training loss on the given dataset; $\bm{T^{admm}}$ is gate-related compression table build on $T$; and $s_i(z_i)$ is an indicator, which will indicate whether the parameter $\theta_i$ will be pruned or not.

\begin{figure}[t]
\centering
\includegraphics[width=1\linewidth]{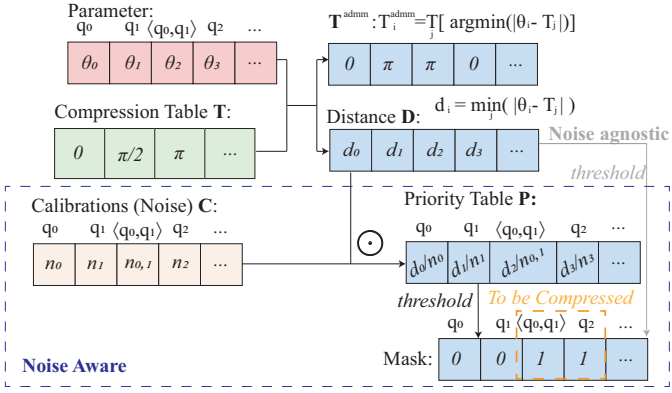}
\caption{Noise-aware mask generation in ADMM process. }
\label{fig:maskexample}
\end{figure} 

In the $r^{th}$ iteration of ADMM optimization, one key step is to determine whether or not to compress a parameter.
According to the parameter $\theta_i$, compression table $T$, and noise data $n_{k,l}$, we will build a mask.
Fig. \ref{fig:maskexample} illustrates the process to create the mask, which is composed of three steps.
First, by comparing the parameter $\theta_i$ with each compression level in table $T$, we generate two tables: $\bm{T^{admm}}$ and $\bm{D}$.
The $i^{th}$ element in $\bm{T^{admm}}$ is denoted as $T_{i}^{admm}$, which is the nearest compression level of parameter $\theta_i$; while $d_i$ is the minimum distance between parameter $\theta_i$ and any compression level.
\textit{Note that in a noise-agnostic compression \cite{hu2022quantum}, a mask will be generated by using table $\bm{D}$.
In the second step, we further consider gate noise and generate a priority table $\bm{P}$.}
Notation $p_i\in \bm{P}$ indicates the priority of gate $g_i$ to be pruned; then, we have $p_i=\frac{C(A(g_i))}{d_i}$, where $A(g_i)$ is the qubits associated with $g_i$ and $C(A(g_i))$ is the noise rate.
Based on $p_i$, we formulate the mask $mask(g_i,\bm{C})^r$ for gate $g_i$ on the given calibration data $\bm{C}$ in the $r^{th}$ iteration as below.
\begin{equation}
\begin{aligned}
mask(g_i ,\bm{C})^r =\begin{cases}
 0& \text{ if } p_i^r < threshold \\ 
 1& \text{ if } otherwise. 
\end{cases}
\end{aligned}
\label{eq:mask}
\end{equation}

In the above formula, the mask equals $1$ indicating that the gate $g_i$ has a high priority, which is larger than a pre-set threshold, to be compressed. To maintain high accuracy, we will utilize the compression level $T_i^{admm}$ if $g_i$ is masked to be compressed.Based on these understandings, we define the indicator function $s_i(z_i)$:
\begin{equation}
\begin{aligned}
s_i(z_i)=\begin{cases}
 0& \text{ if } z_i = T^{admm}_i\times mask(g_i,\bm{C})^r \ \\
  & \ \ \ \ or \ mask(g_i,\bm{C})^r=0 \\ 
 +\infty & \text{ if } otherwise. 
\end{cases}
\end{aligned}
\label{eq:indicator_func}
\end{equation}
Note that $s_i$ is used in Eq. \ref{eq:admm_train} to restrict the value of $z_i$. It requires its value to be 0 for a valid solution.
In the above formula, we set $s_i(z_i)=0$ in two cases: (1) if $mask(g_i ,\bm{C})^r=0$, indicating we do not require the compression on gate $g_i$; or (2) $\theta_i = T^{admm}_i\times mask(g_i,\bm{C})^r$, indicating that the parameter $\theta_i$ has to be compressed to be the compression level $T^{admm}_i$.

For each round, we will get $\bm{\theta}^r$ and $\bm{Z}^r$ alternately. At last, we will get the optimized parameters $\bm{\theta}$ to minimize Eq.~\ref{eq:optprob}.
Then, we will employ noise injection to fine-tune parameters $\bm{\theta}$ to further improve the performance, where we will freeze the compressed parameters to not be tuned using the final $mask$.
\begin{table*}[t]
\centering
\caption{Performance comparison of different methods on 3 datasets in continus 146 days with fluctuating noise.}
\tabcolsep 6pt
\begin{tabular}{|c|cccccccccc|}
\hline
Dataset &
  Method &
  \begin{tabular}[c]{@{}c@{}}Mean\\  Accuracy\end{tabular} &
  \begin{tabular}[c]{@{}c@{}}vs. \\ Baseline\end{tabular} &
  Variance &
  \begin{tabular}[c]{@{}c@{}}Days\\  over 0.8\end{tabular} &
  \begin{tabular}[c]{@{}c@{}}vs.\\  Baseline\end{tabular} &
  \begin{tabular}[c]{@{}c@{}}Days \\ over 0.7\end{tabular} &
  \begin{tabular}[c]{@{}c@{}}vs. \\ Baseline\end{tabular} &
  \begin{tabular}[c]{@{}c@{}}Days \\ over 0.5\end{tabular} &
  \begin{tabular}[c]{@{}c@{}}vs.\\  Baseline\end{tabular} \\ \hline
\multirow{6}{*}{\begin{tabular}[c]{@{}c@{}}4-class \\ MNIST\end{tabular}} &
  Baseline &
  59.35\% &
  0.00\% &
  0.070 &
  24 &
  0 &
  93 &
  0 &
  100 &
  0 \\
 &
  Noise-aware Train Once \cite{wang2022quantumnat} &
  58.69\% &
  -0.65\% &
  0.060 &
  8 &
  -16 &
  92 &
  -1 &
  100 &
  0 \\
 &
  Noise-aware Train Everyday &
  59.39\% &
  0.05\% &
  0.070 &
  28 &
  4 &
  83 &
  -10 &
  99 &
  -1 \\
 &
  One-time Compression \cite{hu2022quantum} &
  68.44\% &
  0.00\% &
  0.050 &
  80 &
  56 &
  102 &
  9 &
  117 &
  17 \\
 &
  QuCAD w/o offline &
  72.31\% &
  12.96\% &
  0.030 &
  77 &
  53 &
  98 &
  5 &
  134 &
  34 \\
 &
  QuCAD (ours) &
  \textbf{75.67\%} &
  \textbf{16.32\%} &
  \textbf{0.020} &
  \textbf{100} &
  \textbf{76} &
  \textbf{134} &
  \textbf{41} &
  \textbf{134} &
  \textbf{34} \\ \hline
\multirow{6}{*}{Iris} &
  Baseline &
  37.85\% &
  0.00\% &
  \textbf{0.006} &
  0 &
  0 &
  0 &
  0 &
  8 &
  0 \\
 &
  Noise-aware Train Once \cite{wang2022quantumnat} &
  54.38\% &
  16.53\% &
  0.043 &
  29 &
  29 &
  46 &
  46 &
  70 &
  62 \\
 &
  Noise-aware Train Everyday &
  56.62\% &
  18.78\% &
  0.044 &
  38 &
  38 &
  56 &
  56 &
  72 &
  64 \\
 &
  One-time Compression \cite{hu2022quantum} &
  69.20\% &
  31.36\% &
  0.043 &
  84 &
  84 &
  90 &
  90 &
  103 &
  95 \\
 &
  QuCAD w/o offline &
  75.30\% &
  37.46\% &
  0.025 &
  \textbf{84} &
  \textbf{84} &
  104 &
  104 &
  128 &
  120 \\
 &
  QuCAD (ours) &
  \textbf{76.73\%} &
  \textbf{38.88\%} &
  0.015 &
  83 &
  83 &
  \textbf{108} &
  \textbf{108} &
  \textbf{141} &
  \textbf{133} \\ \hline
\multirow{6}{*}{\begin{tabular}[c]{@{}c@{}}Seismic\\  Wave\end{tabular}} &
  Baseline &
  68.40\% &
  0.00\% &
  0.014 &
  18 &
  0 &
  70 &
  0 &
  137 &
  0 \\
 &
  Noise-aware Train Once \cite{wang2022quantumnat} &
  68.85\% &
  0.45\% &
  0.014 &
  19 &
  1 &
  78 &
  8 &
  137 &
  0 \\
 &
  Noise-aware Train Everyday &
  68.28\% &
  -0.11\% &
  0.013 &
  22 &
  4 &
  69 &
  -1 &
  138 &
  1 \\
 &
  One-time Compression \cite{hu2022quantum} &
  78.99\% &
  10.59\% &
  0.007 &
  80 &
  62 &
  130 &
  60 &
  144 &
  7 \\
 &
  QuCAD w/o offline &
  82.34\% &
  13.95\% &
  0.001 &
  110 &
  92 &
  145 &
  75 &
  146 &
  9 \\
 &
  QuCAD (ours) &
  \textbf{83.75\%} &
  \textbf{15.36\%} &
  \textbf{0.001} &
  \textbf{133} &
  \textbf{115} &
  \textbf{146} &
  \textbf{76} &
  \textbf{146} &
  \textbf{9} \\ \hline
\end{tabular}
\label{tab:main_result}
\end{table*}

\subsection{Offline Model Repository Constructor}

As discussed in \textit{Motivation 3}, it is possible to use the noise-aware compression to perform optimization before using the QNN, but it is too costly and there exist opportunities to improve efficiency by building a model repository so that the model can be reused.
In this subsection, we will explain how to build the repository using a modified noise-aware k-means clustering algorithm.

There are two inputs of the model repository constructor: (1) offline calibration data, denoted as $\mathbf{C = [c_1,c_2,...,c_n]} \in R^{n\times d}$; and (2) the corresponding QNN accuracy under the calibrations, denoted as $\mathbf{p} = [p_1,p_2,...,p_n] \in R^n$, where $n$ is the number of calibration data and $d$ is the total number of noise rates in each calibration data. 
Our objective is to split calibration data ($\bf{C}$) into $k$ groups, and the samples in the same group are with similar calibration data and performance; meanwhile, the centroid $\bm{r_i}$ is representative of each group ($\bm{g_i}$).
Then, we can use the representative centroid $\bm{r_i}$ to do noise-aware compression and the compressed model will be added to the repository.



\textit{Objective Function and Distance.}
To achieve our goal, we first define performance-aware weight, which is $\mathbf{w}=[w_1,w_2,...,w_d]$, where $w_j$ is the absolute correlation coefficient $\rho = |\frac{\text{cov}(X,Y)}{\sigma_x \sigma_y}|$ between the model performance $\mathbf{p}$ and the $j^{th}$ dimension in calibration data $\mathbf{C_{:,j}}$.
The weighted distance (noted as $ dist_{L1}^w $) between two samples ($\mathbf{c_i}$ and $\mathbf{c_j}$)of calibration data can be defined as
\begin{equation}
dist_{L1}^w(\mathbf{c_i},\mathbf{c_j}) = dist_{L_1}(\mathbf{w \cdot c_i}, \mathbf{w \cdot c_j})
\label{eq:distance}
\end{equation}
where $dist_{L_1}$ is the Manhattan distance between two vectors;

The objective function is designed to partition the data into $K$ clusters such that the sum of weighted Manhattan ($L_1$) absolute errors (WSAE) of all their clusters is minimized. Therefore, the objective function WSAE is designed as:
\begin{equation}
WSAE =  \sum_{\mathbf{g_i} \in \mathbf{G}} \sum_{\forall\mathbf{c} \in \mathbf{g_i}} dist_{L_1}^w(\mathbf{r_i},\mathbf{c})
\label{eq:cluster}
\end{equation}
 $\bf{G}$ is all groups, $\bm{r_i}$ is the representative of group $\bf{g_i}\in\bf{G}$, and $\bm{c}\in \bm{g_i}$ is the candidate calibration in group $\bm{g_i}$.
Since the noise is considered in model performance, the clustering not only considers the value of noise but also the effects of noise on the given QNN model.




\subsection{Online Model Repository Manager}

After building the model repository, the next question is how to use it for online algorithms.
We provide the following guidance to use the repository efficiently.

\textit{Guidance 1:} The cluster results can help to judge whether to generate new models into the model repository manager. At the offline stage, we can get the average weighted distance: $$\overline{(dist_{L1}^w})_{i}= \frac{\sum_{\forall\mathbf{c} \in \mathbf{g_i}} dist_{L_1}(\mathbf{r_i},\mathbf{c})}{n_i} $$ between the centroid($\mathbf{r_i}$) and all samples ($\forall\mathbf{c} \in \mathbf{g_i}$) in the $i^{th}$ cluster, where $n_i$ is the number of samples in the $i^{th}$ cluster. We set $max_{i}(\overline{(dist_{L1}^w})_{i})$ as a threshold $th_w$ to decide whether to add a new centroid (i.e., new representative) in the model repository. If the $min_{j}(dis_j) > th_w$ where $dis_j$ is the $dist_{L1}^w$ distance between the $j^{th}$ centroid and the current calibration data, we will add the current calibration data to the model repository.

\textit{Guidance 2:} The cluster results can be utilized to predict the performance of the given model with current calibration data. 
Specifically, we can obtain the average accuracy of each cluster, say $\overline{acc_i}$ for the $i^{th}$ cluster.
According to users' requirement on QNN model accuracy $A$, we can set cluster $g_i$ as an invalid cluster if its average accuracy $\Bar{acc_i}$ is less than $A$.
Then, at the online stage, if the current calibration matches the centroid in an invalid cluster, we will set the current calibration data as an invalid data, and output a failure report.




 
\section{Experiments}




\subsection{Experiments Setup}

\textbf{Datasets and model.} We evaluate our framework on three classification tasks. (1) We extract 4 class (0,1,3,6) from MNIST \cite{lecun1998gradient} with the former 90\% samples for training and latter 200 samples for testing. To process MNIST data, we apply angle encoding \cite{larose2020robust} to encode $4 \times 4$ images to 4 qubits, and adopt 2 repeats of a VQC block (4RY +4CRY + 4RY +4RX +4CRX +4RX + 4RZ + 4CRZ +4RZ + 4CRZ) as the original model.
(2) We extract 1500 samples of the earthquake detection dataset from FDSN  \cite{fsdn}. Each sample has a positive or negative label. We utilize 90\% and 10\% samples for training and testing, respectively. We encode features to 4 qubits and employ the same VQC as MNIST. 
(3) We use Iris \cite{hoey2004statistical} dataset with 66.6\% and 33.4\% samples for training and testing. Features are encoded to 4 qubits with 3 repeats of VQC blocks.


\textbf{Calibrations and Environment Settings.}
We pull history calibrations from Aug. 10, 2021 to  Sep. 20, 2022 from IBM backend (ibm\_belem) using Qiskit Interface. The front 243 days are used for offline optimization and the remaining 146 days' data are used for online tests. We generate noise models from history calibration data and integrate them into Qiskit noise simulator for online simulation. Besides, we also deploy QuCAD on ibm-jakarta and evaluate the output model of QuCAD on a real IBM quantum processor, ibm-jakarta. Our codes are based on Qiskit APIs and Torch-Quantum~\cite{hanruiwang2022quantumnas}.

\textbf{Competitors.}
We employ different approaches for comparison, including: 
(1) Baseline: training in a noise-free environment without optimization.
(2) Noise-aware Training Once \cite{wang2022quantumnat}: applying noise injection on the first day for training. 
(3) Noise-aware Training Everyday: extend the noise-aware training everyday.
(4) One-time Compression\cite{hu2022quantum}: applying compression with the objective of minimizing circuit length on the first day. 
(5) QuCAD w/o offline: generating the models by our framework without an offline stage.
(6) QuCAD: generating the models by our framework during the offline and online stages.

\subsection{Main Results of our proposed QuCAD}

\textbf{Effective evaluation on noisy simulation.}
Table~\ref{tab:main_result} reports the comparison results of different methods on MNIST, Iris, and earthquake detection datasets. 
Columns ``Mean accuracy'' and ``Variance'' gives the statistical information of model accuracy on the continuously 146 days.
Column ``Days over 0.8'' stands for the number of days that the model accuracy is higher than 80\%, which is similar to ``Days over 0.7'' and ``Days over 0.5'' columns.
From this table, we can clearly see that our proposed QuCAD outperforms all competitors. 
Specifically, on these 3 datasets, QuCAD achieves improvements of 16.32\%, 38.88\%, and 15.36\% respectively, compared with the baseline.
Although there exists one case that the `days over 0.8' of `QuCAD w/o offline' is 1 day more than that of `QuCAD' on Iris, QuCAD has many
days to have accuracy over 70\%.
Besides, QuCAD has the lowest variance except for the baseline on Iris, showing the stability of QuCAD. 
Kindly note that the mean accuracy of the baseline on Iris is much lower than QuCAD.
We also observed that QuCAD outperforms competitors on the number of days over different accuracy requirements, which again demonstrates the effectiveness of our framework.

Compared with noise-aware training, even one-time compression has a significant improvement on all datasets, which validates our observation in Fig. \ref{fig:loss_3d}.
On the other hand, QuCAD w/o offline outperforms one-time compression, indicating that online adaptation is needed. 
Furthermore, compared with `QuCAD w/o offline', QuCAD achieves improvements of 3.36\%, 1.14\%, and 1.41\%, showing the effectiveness of offline optimization in QuCAD.

\begin{figure}[t]
\centering
\includegraphics[width=1.0\linewidth]{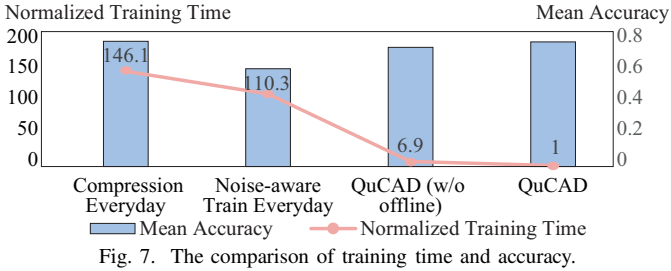}
\caption{The comparison of training time and accuracy.}
\label{fig:effeciency}
\end{figure}
\textbf{Efficiency evaluation.}
We recorded the training time at the online stage in Fig. \ref{fig:effeciency} on 4-class MNIST. 
In the figure, the bars represent the mean accuracy (i.e., right axis) and each point with a number corresponds to a normalized training time (i.e., left axis).
From the results, we can see that QuCAD can achieve 146$\times$ and 110.3$\times$ speedup over ``compression everyday'' and ``noise-aware train every data'', respectively.
This shows the efficiency of our framework, which mainly comes from the reduction of the number of online optimization.
More specifically, the centroids generated offline can well represent the calibrations at the online stage if the distribution of calibrations doesn't change much.

\begin{figure}[t]
\centering
\includegraphics[width=1\linewidth]{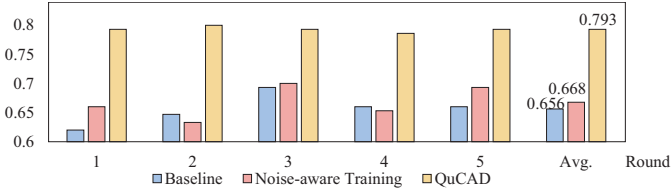}
\caption{On earthquake detection dataset, the performance of different approaches on the 7-qubit quantum device, ibm-jakarta.}
\label{fig:real_device}
\end{figure}
\textbf{Evaluation on real quantum computers.} We also evaluate QuCAD on the earthquake detection dataset using IBM quantum processor ibm-jakarta with different calibration data at different time.
Results are shown in Fig.~\ref{fig:real_device}. 
QuCAD can consistently outperform the two competitors by 13.7\% and 12.52\% on average.
Besides, we can see that the accuracy of QuCAD on different days is more stable than others, which reflects our methods can adapt QNN models to fluctuating noise.

\subsection{Ablation Study}

\begin{table}[t]
\caption{ Comparison of different cluster}
\tabcolsep 5pt
\centering
\begin{tabular}{|c|ccc|}
\hline
Method &
  K &
  \begin{tabular}[c]{@{}c@{}}Mean Acc.\\  of Clusters\end{tabular} &
  \begin{tabular}[c]{@{}c@{}}Mean Acc. \\ of Samples\end{tabular} \\ \hline
K-Means  with L2          & 6           & 72.94\%          & 78.45\%          \\ \hline
Proposed K-Means with $dist_{L1}^w$  & 6         & \textbf{75.83\%} & \textbf{80.68\%} \\ \hline
\end{tabular}
\label{tab:cluster}
\end{table}
\begin{figure}[t]
\centering
\includegraphics[width=1\linewidth]{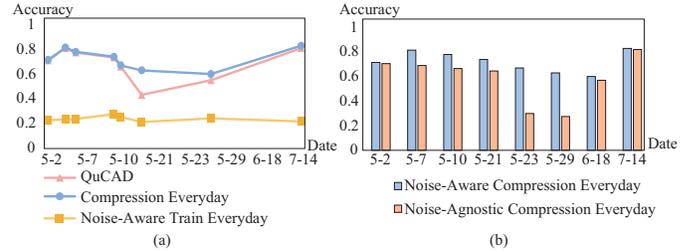}
\caption{Ablation Study: (a) Compression Everyday vs. QuCAD. (b)Noise-Aware vs. Noise-Aanostic training }
\label{fig:ablation10days}
\end{figure} 
\textbf{QuCAD vs. Practical Upper Bound.}
We further investigate the performance of  8 representative days in Fig.~\ref{fig:ablation10days}(a). We apply the result of noise-aware compression every day as a practical upper bound. 
From the results, we can observe that QuCAD has a competitive performance compared to the practical upper bound, showing QuCAD can get approximate optimal results. 

\textbf{Noise-Aware vs. Noise-Agnostic Compression.}
We further evaluate the proposed compression algorithm on those 8 representative days, as shown in Fig.~\ref{fig:ablation10days}(b). Results show that noise-aware compression can outperform noise-agnostic compression on most days, showing the effectiveness of our proposed noise-aware compression. 
We also observe both compression approaches obtained the same accuracy on 5/4 and 7/14. There are two  possible reasons: (1) there is no great difference among qubits; and (2) The noise level is small and the simple compression is enough to get a good performance.

\textbf{Model repository constructor.}
We also did an ablation study on the clustering algorithm developed in the model repository constructor.
We compare our proposed noise-aware distance $dist_{L1}^w$and the standard 'L2 norm' distance.
Results reported in Table~\ref{tab:cluster} show that our method can get 2.89\% higher mean accuracy of 6 clusters on average and obtain 2.23\% accuracy gain on that for all samples consistently.
Results indicate that our proposed method can improve the quality of the centroids and make centroids represent other samples better in a cluster.
\section{Conclusion}
In this work, we reveal the fluctuating noise in quantum computing and it will significantly affect the performance of quantum applications.
To battle against fluctuating noise, we further observe that the high noise level may create breakpoints in the loss surface, and in turn, the noise-aware training may find inferior solutions.
By investigating the breakpoints, we observe that quantum neural network compression can be a hammer to the noise issue.
And we build a compression-aided framework, namely 
QuCAD, which can automatically adapt a given model to fluctuating quantum noise. 
Evaluations on MNIST, earthquake detection dataset, and Iris show the effectiveness and efficiency of QuCAD.
specifically, QuCAD can obtain stable performance on the earthquake detection task using a real IBM quantum processor.



\bibliographystyle{ieeetr}
{\scriptsize
\bibliography{bibliography.bib}
\scriptsize
}

\end{document}